\documentclass[prd,aps,11pt,nofootinbib]{revtex4}
\usepackage{epsfig,color,amssymb,amsmath}
\def\nn{\nonumber}

\def\slash{\!\!\!/}
\def\la{\langle}
\def\ra{\rangle}
\def\r{\gamma}
\def\u{\mu}
\def\v{\nu}
\def\p{\rho}
\def\o{\sigma}
\def\e{\epsilon}
\def\w{\omega}
\def\a{\alpha}
\def\b{\beta}
\def\lmd{\lambda}
\begin{document}
\title{Perturbative QCD Analysis of Exclusive Processes $e^+e^-\rightarrow VP$ and $e^+e^-\rightarrow TP$  }
	
\author{Cai-Dian L\"u$^{1}$~\footnote{Email:lucd@ihep.ac.cn}, Wei Wang$^{2}$~\footnote{Email:wei.wang@sjtu.edu.cn}, Ye Xing$^{2}$~\footnote{Email:xingye\_guang@sjtu.edu.cn}, and Qi-An Zhang$^{1}$~\footnote{Email:zhangqa@ihep.ac.cn}}
	
\affiliation{$^{1}$ Institute of High Energy Physics, Chinese Academy of Science, Beijing 100049, China,\\
School of Physics, University of Chinese Academy of Sciences, Beijing 100049, China;\\
$^{2}$ INPAC, Shanghai Key Laboratory for Particle Physics and Cosmology, \\
MOE Key Laboratory
for Particle Physics, Astrophysics and Cosmology\\ School of Physics and Astronomy, Shanghai JiaoTong University, Shanghai 200240, China  }

\begin{abstract}
We study  the  $e^+e^-\to  VP$ and $e^+e^-\to  TP$ processes in the perturbative QCD approach based on   $k_T$ factorization, where the $P,V$ and $T$ denotes a light pseudo-scalar, vector and tensor meson, respectively.   We point out in the case of $e^+e^-\to  TP$ transition   due to   charge conjugation invariance, only three channels  are allowed: $e^+e^-\to a_2^{\pm} \pi^\mp$, $e^+e^-\to K_2^{*\pm}   K^\mp$ and the V-spin suppressed $e^+e^-\to K_2^{*0} \bar K^0+\overline K_2^{*0}   K^0 $.  Cross sections of $e^+e^-\to  VP$ and $e^+e^-\to  TP$  at $\sqrt{s}=3.67$ GeV and $\sqrt{s}=10.58$ GeV are calculated and the  invariant mass dependence is found to favor the $1/s^4$ power law. Most of our theoretical  results are consistent with the available  experimental data and  other predictions   can be  tested at the ongoing BESIII and  forthcoming Belle-II experiments.
\end{abstract}

\maketitle

\section{Introduction}

The exclusive processes of $e^+e^-$ annihilating  into two mesons   provide an opportunity to investigate  various  time-like meson form factors. The   form factor dependence on  the collision energy $\sqrt{s}$    sheds light on  the structure of partonic constituents in the hadron~\cite{Lu:2007hr,Braguta:2008hs}.  It means that these processes can   be used to extract the  relevant information  on the structure of hadrons in terms of    fundamental quark and gluon degrees of freedom.
Another   reason to study the $e^+e^-$  process  is its  similarity with   annihilation contributions  in charmless  $B$ decays. In two-body charmless $B$ decays,   annihilation diagrams are power-suppressed. However  it has been observed  that in  quite a few decay modes annihilations are rather important \cite{Lu:2002iv,Li:2003wg,Li:2004ep}. Large annihilation diagrams  will very presumably give considerable  strong phases and as a consequence  sizable CP asymmetries are induced \cite{Keum:2000wi,Lu:2000em}.  This fact has an important impact in the CP violation studies of B meson decays. The $e^+e^-\to VP, TP$ processes,  where the $P,V$, and  $T$ denotes a light pseudo-scalar, vector and tensor meson, respectively, have  the  topology  with annihilation diagrams in $B$ decays,  and thus they can  provide an ideal laboratory to isolate power correction effects.

It is anticipated  that hard  exclusive  processes with  hadrons involve  both perturbative and non-perturbative strong interactions.  Factorization, if it exists, allows one to handle the perturbative and non-perturbative contributions separately.  The   short-distance hard kernels can be  calculated perturbatively.  With the nonperturbative inputs determined from other sources,    hard exclusive  processes   provide an effective way to explore the factorization scheme.
The factorization theorem ensures that a physical amplitude can be expressed as a  convolution of hard scattering kernels and hadron distribution amplitudes.  However if one directly applies the collinear factorization to the  $e^+e^-\to VP,TP$, the amplitude diverges  in the end point region $x\to 0$. Here $x$ is the momentum fraction of the involved quark.

A modified perturbative QCD approach based on $k_T$ factorization,   called PQCD approach for brevity, is proposed~\cite{Li:1992nu,Li:1994cka,Keum:2000ph,Keum:2000wi,Lu:2000em,Lu:2000hj} and has been  successfully applied to many reactions~\cite{Li:2000hh,Nagashima:2002ia,Nagashima:2002iw,Li:2004ep,Liu:2005mm,Wang:2005bk,Shen:2006ms,Ali:2007ff,Zhang:2008by,Li:2008tk,Li:2009tx,Wang:2010ni,Xiao:2011tx,Kim:2013ria,Bai:2013tsa,Zhang:2015vor,Zou:2015iwa}. In this approach,  the transverse momentum of partons in the meson is kept to  kill endpoint divergences.  Then the physical amplitude is  written as a convolution of the universal non-perturbative hadronic wave functions   and  hard   kernels in both longitudinal and transverse directions. Double logarithms, arising from the overlap of the soft and collinear divergence, can be resumed into  Sudakov factor, while single logarithms from ultraviolet divergences can be handled by renormalization group equation (RGE). With  Sudakov factor taken into account, the applicability of perturbative QCD can be brought  down to a few GeV.
In this work, we will study the  $e^+e^-\rightarrow VP$ and $e^+e^-\rightarrow TP$  in the perturbative QCD (PQCD) approach~\cite{Li:1992nu,Li:1994cka,Keum:2000ph,Keum:2000wi,Lu:2000em,Lu:2000hj} based on $k_T$ factorization.

The rest of this paper is organized as follows. In section II, we  first collect the input parameters including  decay constants and light-cone wave functions. Then we present the  PQCD framework  and  give factorization formulas for the time-like form factors. Numerical results and detailed  discussions are presented in section III. The last section contains the conclusion.

\section{Perturbative QCD calculation}

\subsection{Notations}

We consider the $e^+e^-\rightarrow V(T)P$, in which $V(T)$ is a vector (tensor) meson with momentum $P_1$ and polarization vector $\e_{\u}$ (polarization tensor $\e_{\u\v})$, and $P$ denotes a  pseudoscalar meson with momentum $P_2$  in the center of mass frame. The collision energy  is denoted as $Q=\sqrt{s}$. In  the standard model, such processes    proceed through a virtual photon or a $Z^0$ boson. At low energy with $\sqrt {s}\sim $ a few GeV,   the amplitude is dominated by a photon. In this case the hadron amplitude is parameterized  in terms of a   form factor:
\begin{align}
\la V(P_1,\e_T)P(P_2)|j_{\u}^{\mathrm{em}}|0\ra = F_{\mathrm{VP}}(s)\e_{\u\v\a\b}\e_T^{\v}P_1^{\a}P_2^{\b}. \label{VPFF}
\end{align}
Notice that in Eq.(\ref{VPFF}) the vector meson is transversely polarized. We have adopted the convention $\e^{0123}=1$ for the Levi-Civita tensor.

For a tensor meson, its  polarization tensor $\e_{\u\v}$ can be constructed via the polarization vector
\begin{align}
\e_{\u}(0)=\frac{1}{m_T}(|\vec{P}_T|,0,0,E_T),~~~ \e_{\u}(\pm)=\frac{1}{\sqrt{2}}(0,\mp1,-i,0).
\end{align}
Using the  Clebsch-Gordan coefficients~\cite{Olive:2016xmw}, one has
\begin{align}
&\e_{\u\v}{(\pm2)}=\e_{\u}{(\pm)}\e_{\v}{(\pm)}, \nn\\
&\e_{\u\v}{(\pm1)}=\sqrt{\frac{1}{2}} \big[\e_{\u}{(\pm)}\e_{\v}(0)+\e_{\u}(0)\e_{\v}{(\pm)}\big], \nn\\
&\e_{\u\v}{(\pm0)}=\sqrt{\frac{1}{6}} \big[\e_{\u}{(+)}\e_{\v}{(-)}+\e_{\u}{(-)}\e_{\v}{(+)}\big]+\sqrt{\frac{2}{3}} \e_{\u}(0)\e_{\v}(0). \label{CGC}
\end{align}
In the calculation it is convenient to introduce a new polarization vector $\xi$:
\begin{align}
\xi_{\u}(\lmd)=\frac{\e_{\u\v}(\lmd)q^{\v}}{P_1\cdot q}m_T, \label{polaTensor}
\end{align}
where $q=P_1+P_2$ is the four momentum of the virtual photon and $q^2=s$. Then  Eq.(\ref{CGC}) becomes
\begin{align}
&\xi_{\u}(\pm2)=0,~~~\xi_{\u}(\pm1)=\frac{1}{\sqrt{2}}\frac{Q^2\eta}{2m_T^2+Q^2\eta}\e_{\u}(\pm),~~~\xi_{\u}(0)=\sqrt{\frac{2}{3}}\frac{Q^2\eta}{2m_T^2+Q^2\eta}\e_{\u}(0),
\end{align}
where $\eta=1-m_T^2/Q^2$, with $m_T$ as the mass of the tensor meson.  Here the mass of the pseudoscalar meson has  been neglected.  The new    vector $\xi$ plays a similar role with the ordinary polarization vector $\e$, regardless of some  dimensionless constants.

Then like Eq.~\eqref{VPFF}, one  can define the $TP$ form factor  as
\begin{align}
\la T(P_1,\lmd)P(P_2)|j_{\u}^{\mathrm{em}}|0\ra= F_{\mathrm{TP}}\e_{\u\v\a\b}\xi^{\v}(\lmd)P_1^{\a}P_2^{\b}, \label{TPFF}
\end{align}
in which the final state tensor meson   is also transversely  polarized.

Using  the form factors in Eqs.(\ref{VPFF},\ref{TPFF}), one can derive  the cross sections for $e^+e^-\rightarrow VP, TP$
\begin{align}
&\o(e^+e^-\rightarrow VP)=\frac{\pi\a_{\mathrm{em}}^2}{6}|F_\mathrm{VP}|^2\Phi^{3/2}(s), \\
&\o(e^+e^-\rightarrow TP)=\frac{\pi\a_{\mathrm{em}}^2}{3}\Big(\frac{s\eta}{2m_T^2+s\eta}\Big)^2|F_\mathrm{TP}|^2\Phi^{3/2}(s),
\end{align}
with the fine structure constant $\alpha_{\rm em}=1/137$, and
\begin{align}
\Phi(s)=\bigg[1-\frac{(m_{V(T)}+m_P)^2}{s}\bigg]\bigg[1-\frac{(m_{V(T)}-m_P)^2}{s}\bigg].
\end{align}

\subsection{Decay constants and Light cone wave functions}

Decay constants for a pseudoscalar meson and a vector meson   are defined by:
\begin{eqnarray}
\la P(p)|\bar{q}_2\r_{\u}\r_5q_1|0\ra &=&-if_Pp_{\u}, \\
 \la V(p,\e)|\bar{q}_2\r_{\u}q_1|0\ra&=& f_Vm_V\e_{\u}, \;
 \la V(p,\e)|\bar{q}_2\o_{\u\v}q_1|0\ra=-if_V^T(\e_{\u}p_{\v}-\e_{\v}p_{\u}).
\end{eqnarray}
Tensor meson decay constants are defined as \cite{Cheng:2010hn}
\begin{align}
&\la T(P,\lambda)|j_{\u\v}(0)|0\ra=f_Tm_T^2\e_{\u\v}^{(\lambda)*}, \;\;
 \la T(P,\lambda)|j_{\u\v\delta}^{\perp}(0)|0\ra=-if_T^{\perp}m_T(\e_{\u\delta}^{(\lambda)*}P_{\v}-\e_{\v\delta}^{(\lambda)*}P_{\u}).
\end{align}
The interpolating currents are chosen as
\begin{align}
&j_{\u\v}(0)=\frac{1}{2}\Big(\bar{q}_1(0)\r_{\u}i\overleftrightarrow{D}_{\v}q_2(0)+\bar{q}_1(0)\r_{\v}i\overleftrightarrow{D}_{\u}q_2(0)\Big), \\
&j_{\u\v\delta}^{\perp\dagger}(0)=\bar{q}_2(0)\o_{\u\v}i\overleftrightarrow{D}_{\delta}q_1(0),
\end{align}
with the covariant derivative $\overleftrightarrow{D}_{\u}=\overrightarrow{D}_{\u}-\overleftarrow{D}_{\u}$ with $\overrightarrow{D}_{\u}=\overrightarrow{\partial}_{\u}+ig_sA_{\a}^a\lambda^a/2$ and $\overleftarrow{D}_{\u}=\overleftarrow{\partial}_{\u}-ig_sA_{\a}^a\lambda^a/2$.

The pseudoscalar and vector decay constants can be determined from various reactions, $\pi^-\to e^-\bar\nu$, $\tau^-\rightarrow (\pi^-, K^-\p^-,K^{*-})\v_{\tau}$ and $V^0\rightarrow e^+e^-$~\cite{Olive:2016xmw}. For tensor mesons, their decay constants can be  calculated in   QCD sum rules \cite{Aliev:1981ju,Aliev:2009nn} and we quote the recently updated results from Ref. \cite{Cheng:2010hn}. Results for decay constants   are collected in Table \ref{tab:DecayConstant}.

\begin{table}[htbp]
\caption{Decay constants of the relevant light mesons (in units of MeV)}\label{tab:DecayConstant}
\begin{center}
\scriptsize\begin{tabular}{cccccccccccccc}
\hline\hline
 $f_{\pi}$ & $f_K$ & $f_{\p}$ & $f_{\p}^T$ & $f_{\w}$ & $f_{\w}^T$ & $f_{K^*}$ & $f_{K^*}^T$ & $f_{\phi}$ & $f_{\phi}^T$ & $f_{a_2}$ & $f_{a_2}^T$ & $f_{K_2^*}$ & $f_{K_2^*}^T$ \\
 $131$ & $160$ & $209\pm 2$ & $165\pm9$ & $195\pm3$ & $145\pm10$ & $217\pm5$ & $185\pm10$ & $231\pm4$ & $200\pm10$ & $107\pm6$ & $105\pm21$ & $118\pm5$ & $77\pm14$ \\
\hline\hline
\end{tabular}
\end{center}
\end{table}

The light-cone distribution amplitudes (LCDAs) are defined as   matrix elements of   non-local operators at the light-like separations $z_{\u}$ with $z^2=0$, and sandwiched between the vacuum and the meson state. The two-particle LCDAs of a pseudoscalar meson, up to twist-3 accuracy, are defined by \cite{Ball:1998je}
\begin{align}
\la P(p)|\bar{q}_{2\b}(z)q_{1\a}(0)|0\ra &= \frac{-i}{\sqrt{2N_C}}\int_0^1dxe^{ixp\cdot z}\Big[\r_5p\slash\phi_P^A(x)+m_0\r_5\phi_P^P(x)+m_0\r_5(n\slash v\slash-1)\phi_P^T(x)\Big]_{\a\b},
\end{align}
where $n$, $v$ are two light-like vectors. The final-state P meson is moving on the $n$ direction with $v$ the opposite direction.  $x$ is the momentum fraction carried by the quark $q_2$. The chiral enhancement parameter $m_0= {m_P^2}/({m_{q_1}+m_{q_2}})$, is used in our work as    $m_0^{\pi}=1.4\pm0.1$GeV, $m_0^K=1.6\pm0.1$GeV \cite{Ball:2004ye,Ball:2006wn}.

We use the following form for leading twist LCDAs derived from the conformal symmetry:
\begin{align}
\phi_P^A(x)=\frac{3f_P}{\sqrt{2N_C}}x(1-x)[1+a_1^PC_1^{3/2}(t)+a_2^PC_2^{3/2}(t)], \label{phiPA}
\end{align}
where $N_C=3$ and  $t=2x-1$. $C_i^{3/2} (i=1,2)$ are Gegenbauer polynomials, with the definition
\begin{align}
C_1^{3/2}(t)=3t,~~~C_2^{3/2}(t)=\frac{3}{2}(5t^2-1).
\end{align}
The Gegenbauer moments at $\u=1$GeV are used  as \cite{Ball:2004ye,Ball:2006wn}:
\begin{align}
&a_1^{\pi}=0,~~~a_1^K=0.06\pm0.03,~~~a_2^{\pi,K}=0.25\pm0.15.
\end{align}

In this paper, we will study the collision at $\sqrt{s}=3.67$GeV and $10.58$GeV, and  then  it is plausible  to adopt the asymptotic forms for twist-3 DAs for simplicity:
\begin{align}
\phi_{P}^P(x)=\frac{f_{P}}{2\sqrt{2N_c}},~~~\phi_{P}^T(x)=\frac{f_{P}}{2\sqrt{2N_C}}(1-2x). \label{phiPP}
\end{align}

As for the $\eta-\eta'$ mixing, we use the quark flavor  basis with the mixing scheme~\cite{Feldmann:1998vh,Feldmann:1998sh}:
\begin{align}
\begin{pmatrix} \eta \\ \eta' \end{pmatrix} = U(\phi)\begin{pmatrix} \eta_q \\ \eta_s \end{pmatrix} = \begin{pmatrix} \cos\phi & -\sin\phi \\ \sin\phi & \cos\phi \end{pmatrix}\begin{pmatrix} \eta_q \\ \eta_s \end{pmatrix}.  \label{etaMixing}
\end{align}
The mixing angle is  $\phi=39.3^{\circ}\pm1.0^{\circ}$\cite{Feldmann:1998vh,Feldmann:1998sh} and
\begin{align}
\eta_q=\frac{1}{\sqrt{2}}(u\bar{u}+d\bar{d}),~~~\eta_s=s\bar{s}.
\end{align}
Their decay constants are defined as:
\begin{align}
\la0|\bar{n}\r^{\u}\r_5n|\eta_n(P)\ra=\frac{i}{\sqrt{2}}f_nP^{\u},~~~
\la0|\bar{s}\r^{\u}\r_5s|\eta_s(P)\ra=if_sP^{\u}.
\end{align}
In the following calculation, we  will assume the same  wave functions for the $n\bar{n}$ and $s\bar{s}$  as the pion's wave function,  except for the different decay constants \cite{Feldmann:1998vh,Feldmann:1998sh} and the chiral scale parameters \cite{Charng:2006zj}:
\begin{align}
f_n=(1.07\pm0.02)f_{\pi},~~~f_s=(1.34\pm0.06)f_{\pi},~~~m_0^n=1.07\mathrm{GeV},~~~m_0^s=1.92\mathrm{GeV}.
\end{align}

Similar  with   pseudoscalar mesons, the two-particle LCDAs for transversely polarized vector mesons up to twist-3  are parameterized as \cite{Ball:1998sk,Ball:1998ff}:
\begin{align}
\la V(p,\e_T)|\bar{q}_{2\b}(z)q_{1\a}(0)|0\ra =\frac{1}{\sqrt{2N_C}}\int_0^1dxe^{ixp\cdot z}\Big[&\e\slash_Tp\slash\phi_V^T(x)+m_V\e\slash_T\phi_V^v(x) \nn\\
+&m_Vi\e_{\u\v\p\o}\r_5\r^{\u}\e_T^{\v}n^{\p}v^{\o}\phi_V^p(x)\Big]_{\a\b}.
\end{align}
The twist-2 LCDA  can be expanded as:
\begin{align}
\phi_V^T(x)=\frac{3f_V^T}{\sqrt{2N_C}}x(1-x)[1+a_1^{\perp}C_1^{3/2}(t)+a_2^{\perp}C_2^{3/2}(t)],
\end{align}
with  Gegenbauer moments at $\u=1$GeV~\cite{Ball:2004rg,Ball:2006nr}:
\begin{align}
&a_{1K^*}^{\perp}=0.04\pm0.03,~~~a_{1\p}^{\perp}=a_{1\w}^{\perp}=a_{1\phi}^{\perp}=0, \nn\\
&a_{2K^*}^{\perp}=0.11\pm0.09,~~~a_{2\p}^{\perp}=a_{2\w}^{\perp}=0.15\pm0.07,~~~a_{2\phi}^{\perp}=0.06_{-0.07}^{+0.09}.
\end{align}

As for the twist-3 LCDAs,  we will also use the asymptotic forms:
\begin{align}
&\phi_V^v(x)=\frac{3f_V}{8\sqrt{2N_C}}[1+(2x-1)^2], \;\;
\phi_V^p(x)=\frac{3f_V}{4\sqrt{2N_C}}(1-2x).
\end{align}

For a generic tensor meson,   the LCDAs up to twist-3 can be defined  as \cite{Wang:2010ni}:
\begin{eqnarray}
\la T(p,\pm1)|\bar{q}_{2\b}(z)q_{1\a}(0)|0\ra &=& \frac{1}{\sqrt{2N_C}}\int_0^1e^{ixp\cdot z} \nn\\
&&\times \Big[\xi\slash_Tp\slash\phi_T^T(x) +m_T\xi\slash_T\phi_T^V(x)+m_Ti\e_{\u\v\p\o}\r_5\r^{\u}\xi_T^{\v}n^{\p}v^{\o}\phi_T^a(x)\Big]_{\a\b}.  \end{eqnarray}
These LCDAs are related to the ones given in  \cite{Cheng:2010hn}:
\begin{align}
\phi_T^T(x)=\frac{f_T^T}{2\sqrt{2N_C}}\phi_{\perp}(x),~~~\phi_T^V(x)=\frac{f_T}{2\sqrt{2N_C}}g_{\perp}^{(v)}(x),~~~\phi_T^a(x)=\frac{f_T}{8\sqrt{2N_C}}\frac{d}{dx}g_{\perp}^{(a)}(x).
\end{align}
The asymptotic forms will be  used in the calculation:
\begin{eqnarray}
\phi_{\parallel,\perp}(x)=30x(1-x)(2x-1),\\
g_{\perp}^{(a)}(x)=20x(1-x)(2x-1),~~~g_{\perp}^{(v)}(x)=5(2x-1)^3.
\end{eqnarray}

In the above, we have only discussed  the longitudinal momentum distributions.   It is reasonable  that the  transverse momentum also plays an important role. Thus  we will include the transverse momentum dependent parton distributions (TMDs) of the final-state light mesons. Following Ref. \cite{Lu:2007hr}, we assume  no interference between the longitudinal and transverse   distributions, and thus one can use the  following Gaussian forms  to factorize the wave functions~\cite{Stefanis:1998dg,Kurimoto:2006iv}:
\begin{align}
&\psi(x,\mathbf{b})=\phi(x)\times\exp\Big(-\frac{b^2}{4\b^2}\Big), \label{wf1}\\
&\psi(x,\mathbf{b})=\phi(x)\times\exp\Big[-\frac{x(1-x)b^2}{4a^2}\Big]. \label{wf2}
\end{align}
In the above equation  $\phi(x)$ is the longitudinal momentum distribution amplitude, and the exponential factor describes the transverse momentum distribution. The parameters $\b$ and $a$ characterize the shape of the transverse momentum distributions. The parameter $\b$ is expected at the order of $\Lambda_{\mathrm{QCD}}$ and  related with the  root  of  the averaged  transverse momentum square $\la\mathbf{k}_T^2\ra^{1/2}$. If we choose $\la\mathbf{k}_T^2\ra^{1/2}=0.35$GeV,  $\b^2=4{\rm GeV}^{-2}$. According to Ref.~\cite{Kurimoto:2006iv}, the   size parameter $a$ follows $a^{-1}\simeq\sqrt{8}\pi f_M$, where $f_M$ is the decay constant of the related hadron.

\subsection{PQCD Calculation}

In the PQCD scheme, a form factor can be written as the convolution of a hard scattering kernel with universal hadron wave functions. In small-$x$ region, the parton transverse momentum $k_T$ is at the same order   with  the longitudinal momentum. Once $k_T$ is introduced in the hard   kernel, a   transverse momentum dependent (TMD) wave function is requested. Then the form factor is factorized as:
\begin{eqnarray}
F(Q^2)&=&\int_0^1dx_1dx_2\int d^2\mathbf{k_{T1}}d^2\mathbf{k_{T2}}
\Phi_{M_1}(x_1,\mathbf{k_{T1}},P_1,\u)H(x_1,x_2,\mathbf{k_{T1}},\mathbf{k_{T2}},Q,\u)\Phi_{M_2}(x_2,\mathbf{k_{T2}},P_2,\u) \nonumber\\
&=&\int_0^1dx_1dx_2\int\frac{d^2\mathbf{b_1}}{(2\pi)^2}\frac{d^2\mathbf{b_2}}{(2\pi)^2}
\mathcal{P}_{M_1}(x_1,\mathbf{b_1},P_1,\u)H(x_1,x_2,\mathbf{b_1},\mathbf{b_2},Q,\u)\mathcal{P}_{M_2}(x_2,\mathbf{b_2},P_2,\u).\label{FourierTrans}
\end{eqnarray}
Eq.(\ref{FourierTrans}) is the  Fourier form in the impact parameter $b$ space. Here $\Phi_{M_i}(x_i,\mathbf{k}_{Ti},P_i,\u)$ and $\mathcal{P}_{M_i}(x_i,\mathbf{b}_i,P_i,\u)$ are both the hadron wave functions, relying on $k_T$ and $b$ respectively.

Double logarithms arising from the overlap of soft and collinear divergences, can be resumed into   Sudakov factor \cite{Botts:1989kf,Stefanis:2000vd}:
\begin{align}
\mathcal{P}_{M_i}(x_i,\mathbf{b}_i,P_i,\u)=\exp[-s(x_i,b_i,Q)-s(1-x_i,b_i,Q)]\mathcal{P}_{M_i}(x_i,\mathbf{b}_i,\u).\label{111}
\end{align}
The Sudakov factor $s(\xi,b_i,Q),\xi=x_i$ or $1-x_i$, is given as \cite{Cao:1995eq,Li:1997un}:
\begin{align}
s(\xi,b,Q)~=~&\frac{A^{(1)}}{2\b_1}\hat{q}\ln\Big(\frac{\hat{q}}{\hat{b}}\Big)+\frac{A^{(2)}}{4\b_1^2}\Big(\frac{\hat{q}}{\hat{b}}-1\Big)-\frac{A^{(1)}}{2\b_1}\Big(\hat{q}-\hat{b}\Big)
-\frac{A^{(1)}\b_2}{4\b_1^3}\hat{q}\bigg[\frac{\ln(2\hat{b})+1}{\hat{b}}-\frac{\ln(2\hat{q})+1}{\hat{q}}\bigg] \nn\\
&-\bigg[\frac{A^{(2)}}{4\b_1^2}-\frac{A^{(1)}}{4\b_1}\ln\Big(\frac{e^{2\r-1}}{2}\Big)\bigg]\ln\Big(\frac{\hat{q}}{\hat{b}}\Big)+\frac{A^{(1)}\b_2}{8\b_1^3}\big[\ln^2(2\hat{q})-\ln^2(2\hat{b})\big],
\end{align}
where  the  notations have been used:
\begin{align}
&\hat{q}\equiv\ln\Big[\frac{\xi Q}{\sqrt{2}\Lambda_{\mathrm{QCD}}}\Big],~~~\hat{b}\equiv\ln\Big[\frac{1}{b\Lambda_{\mathrm{QCD}}}\Big].
\end{align}
The  running coupling constant is given as
\begin{align}
\frac{\a_s}{\pi}=\frac{1}{\b_1\log(\u^2/\Lambda_\mathrm{QCD}^2)}-\frac{\b_2}{\b_1^3}\frac{\ln\ln(\u^2/\Lambda_\mathrm{QCD}^2)}{\ln^2(\u^2/\Lambda_\mathrm{QCD}^2)},
\end{align}
and the coefficients $A^{(i)}$ and $\b_i$ are
\begin{align}
&\b_1=\frac{33-2n_f}{12},~~~\b_2=\frac{153-19n_f}{24},\nn\\
&A^{(1)}=\frac{4}{3},~~~A^{(2)}=\frac{67}{9}-\frac{\pi^2}{3}-\frac{10}{27}n_f+\frac{8}{3}\b_1\ln\Big(\frac{e^{\r_E}}{2}\Big).
\end{align}
Here $n_f$ is the number of the quark flavors and $\r_E$ is the Euler constant.

Apart from the double logarithms,   single logarithms from ultraviolet divergence   emerge in the radiative corrections to both the hadronic wave functions and hard kernels. These are summed by the renormalization group (RG) method:
\begin{align}
\Big[\u\frac{\partial}{\partial\u}+\b(g)\frac{\partial}{\partial g}\Big]\mathcal{P}_{M_i}(x_i,\mathbf{b_i},P_i,\u)&=-2\r_q\mathcal{P}_{M_i}(x_i,\mathbf{b_i},P_i,\u), \\
\Big[\u\frac{\partial}{\partial\u}+\b(g)\frac{\partial}{\partial g}\Big]H(x_1,x_2,\mathbf{b_1},\mathbf{b_2},Q,\u)&=4\r_qH(x_1,x_2,\mathbf{b_1},\mathbf{b_2},Q,\u).
\end{align}
Here the quark anomalous dimension   is $\r_q=-\a_s/\pi$. In terms of the above equations, we can get the RG evolution of the hadronic wave functions and hard scattering amplitude as
\begin{align}
\mathcal{P}_{M_i}(x_i,\mathbf{b_i},P_i,\u)&=\exp\Big[-2\int_{1/b_i}^{\u}\frac{d\bar{\u}}{\bar{\u}}\r_q\big(\a_s(\bar{u})\big)\Big]\times\bar{\mathcal{P}}_{M_i}(x_i,\mathbf{b_i},1/b_i),\label{333}\\
H(x_1,x_2,\mathbf{b_1},\mathbf{b_2},Q,\u)&=\exp\Big[-4\int_{\u}^{t}\frac{d\bar{\u}}{\bar{\u}}\r_q\big(\a_s(\bar{u})\big)\Big]\times H(x_1,x_2,\mathbf{b_1},\mathbf{b_2},Q,t),
\end{align}
where $t$ is the largest energy scale in the hard scattering.
Then from equations (\ref{111}) and (\ref{333}), the large-$b$ behavior of $\mathcal{P}$ can be summarized as
\begin{align}
\mathcal{P}_{M_i}(x_i,\mathbf{b}_i,P_i,\u)=\exp[-S(x_i,b_i,Q,\u)]\mathcal{P}_{M_i}(x_i,\mathbf{b}_i,1/b_i),
\end{align}
with
\begin{align}
S(x_i,b_i,Q,\u)=s(x_i,b_i,Q)+s(1-x_i,b_i,Q)+2\int_{1/b_i}^{\u}\frac{d\bar{\u}}{\bar{\u}}\r_q\big(\a_s(\bar{\u})\big).
\end{align}

Furthermore, QCD loop corrections for the electromagnetic vertex can  induce another type of double logarithms $\a_s\ln^2x_i$. They are usually factorized from the hard amplitude and resummed into the jet function $S_t(x_i)$ to further suppress the end-point contribution. It should be pointed out that   Sudakov factor from threshold resummation is universal and independent on the flavors of internal quarks, twist and topologies of hard scattering amplitudes and the specific process \cite{Sterman:1986fn,Sterman:1986aj,Catani:1989ne,Li:1998is,Li:2001ay}. The following approximate parametrization is proposed in \cite{Kurimoto:2001zj} for the convenience of phenomenological applications
\begin{align}
S_t(x,Q)=\frac{2^{1+2c}\Gamma(3/2+c)}{\sqrt{\pi}\Gamma(1+c)}[x(1-x)]^c,
\end{align}
in which the  $c$ is a parameter depending  on $Q$. Ref.~\cite{Li:2009pr} proposed a parabolic parametrization of the   $Q^2$ dependence:
\begin{align}
c(Q^2)=0.04Q^2-0.51Q+1.87,
\end{align}
The threshold resummation modifies the end point behavior of the hadron wave functions, rendering them vanish faster in this region.

\begin{figure}
\includegraphics[width=0.55\columnwidth]{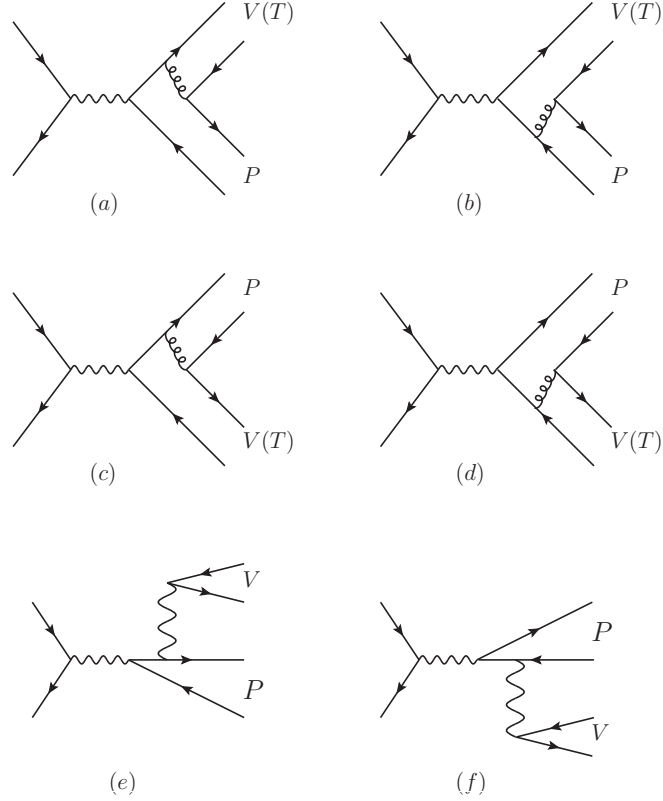}
\caption{Feynman diagrams for $e^+e^-\to VP,TP$.  In the first four panels, a hard momentum transfer occur through the highly virtual gluon. In the last two panels, the neutral vector meson is generated by a photon.  }
\label{fig:feynman}
\end{figure}


Taking   into account all the above ingredients, one  can obtain the analytic results of  the  first four diagrams in Fig.~\ref{fig:feynman} in $k_T$ factorization:
\begin{align}
F_a=&16\pi C_FQ\int_0^1dx_1dx_2\int_0^{\infty}b_1db_1b_2db_2E(t_a)h(\bar{x}_1,x_2,b_1,b_2)S_t(x_2) \nn\\ &\times \Big\{r_1\big[\phi_1^{p(a)}(x_1,b_1)-\phi_1^{v}(x_1,b_1)\big]\phi_2^A(x_2,b_2)\Big\}, \label{FaResult}\\
F_b=&16\pi C_FQ\int_0^1dx_1dx_2\int_0^{\infty}b_1db_1b_2db_2E(t_b)h(x_2,\bar{x}_1,b_2,b_1)S_t(\bar{x}_1) \nn\\ &\times
\Big\{r_1\bar{x}_1\big[\phi_1^{p(a)}(x_1,b_1)+\phi_1^{v}(x_1,b_1)\big]\phi_2^A(x_2,b_2)-2r_2\phi_1^T(x_1,b_1)\phi_2^P(x_2,b_2)\Big\}, \\
F_c=&-16\pi C_FQ\int_0^1dx_1dx_2\int_0^{\infty}b_1db_1b_2db_2E(t_c)h(\bar{x}_2,x_1,b_2,b_1)S_t(x_1) \nn\\ &\times
\Big\{r_1x_1\big[\phi_1^{p(a)}(x_1,b_1)-\phi_1^v(x_1,b_1)\big]\phi_2^A(x_2,b_2)+2r_2\phi_1^T(x_1,b_1)\phi_2^P(x_2,b_2)\Big\}, \\
F_d=&-16\pi C_FQ\int_0^1dx_1dx_2\int_0^{\infty}b_1db_1b_2db_2E(t_d)h(x_1,\bar{x}_2,b_1,b_2)S_t(\bar{x}_2) \nn\\ &\times
\Big\{r_1\big[\phi_1^{p(a)}(x_1,b_1)+\phi_1^v(x_1,b_1)\big]\phi_2^A(x_2,b_2)\Big\}, \label{FdResult}
\end{align}
where $E(t_i)$ and $h$ are  given as
\begin{align}
&E(x_1,x_2,b_1,b_2,Q,t_i)=\a_s(t_i)\exp[-S_1(x_1,b_1,Q,t_i)-S_2(x_2,b_2,Q,t_i)], \\
&h(x_1,x_2,b_1,b_2,Q)=\Big(\frac{i\pi}{2}\Big)^2H_0^{(1)}(\sqrt{x_1x_2}Qb_1)\big[\theta(b_1-b_2)H_0^{(1)}(\sqrt{x_2}Qb_1)J_0(\sqrt{x_2}Qb_2) \nn\\
&~~~~~~~~~~~~~~~~~~~~~~~~~~~~~~~~~~~~~~~~~~~~~~~~~~~+\theta(b_2-b_1)H_0^{(1)}(\sqrt{x_2}Qb_2)J_0(\sqrt{x_2}Qb_1)\big],
\end{align}
where $J_0$ and $H_0^{(1)}$ are both Bessel functions. We take $\bar{x}=1-x$ for short and define $r_i=m_i/Q$, with the index $i=1,2$ for the cases of final state meson is vector(tensor) or pseudoscalar meson. The factorization scale $t$ is chosen as  the largest mass scale involved in the hard scattering:
\begin{align}
&t_a=\max(\sqrt{x_2}Q,1/b_1,1/b_2),~~~t_b=\max(\sqrt{\bar{x}_1}Q,1/b_1,1/b_2), \nn\\
&t_c=\max(\sqrt{x_1}Q,1/b_1,1/b_2),~~~t_d=\max(\sqrt{\bar{x}_2}Q,1/b_1,1/b_2). \label{hst1}
\end{align}

If the final state meson is not  a strange meson, the distribution amplitudes are completely symmetric or antisymmetric under the interchange of the quark and antiquark's momentum fraction $x$ and $1-x$. Then one can   obtain
\begin{align}
&F_a(VP)=F_d(VP),~~~F_b(VP)=F_c(VP); \label{VPrelationsOfABCD}\\
&F_a(TP)=-F_d(TP),~~~F_b(TP)=-F_c(TP). \label{TPrelationsOfABCD}
\end{align}

The contributions from a photon radiated from  the  interaction point into a vector meson, shown as the last two panels  in Fig.~\ref{fig:feynman}, might be sizable. Although these diagrams are suppressed by   $\a_{em}$, they are  enhanced by the  almost on-shell photon propagator ($1/m_V^2$) compared with the gluon propagator in the first four diagrams ($\sim 1/s$)~\cite{Beneke:2005we,Lu:2006nza,Wang:2017gay,Guo:2016fqg}. These two amplitudes can be calculated   in collinear factorization due to the absence of endpoint singularities in these two diagrams. In particular, they are equal after integrating out the momentum fractions:
\begin{align}
F_e=F_f=\frac{12\pi\a_{em}^2f_Pf_{V }}{m_{V}s}(1+a_2^P).
\end{align}

Finally, the form factors for the explicit channels of $e^+e^-\rightarrow VP$ process are   combinations of the six amplitudes $F_{a-f}$:
\begin{align}
&F_{\p^+\pi^-}~=~F_{\p^-\pi^+}~=~\frac{1}{3}\big[ F_a(\p\pi)+F_b(\p\pi) \big], \\
&F_{\p^0\pi^0}~=~\frac{1}{3}\big[ F_a(\p\pi)+F_b(\p\pi) \big] + \frac{1}{6}\big[ F_e(\p\pi)+F_f(\p\pi) \big], \\
&F_{K^{*+}K^-}~=~\frac{2}{3}\big[ F_a(K^*K)+F_b(K^*K) \big] - \frac{1}{3}\big[ F_c(K^*K)+F_d(K^*K) \big], \label{K+K-VP} \\
&F_{K^{*-}K^+}~=~-\frac{1}{3}\big[ F_a(K^*K)+F_b(K^*K) \big] + \frac{2}{3}\big[ F_c(K^*K)+F_d(K^*K) \big], \\
&F_{K^{*0}\bar{K}^0}~=~F_{\bar{K}^{*0}K^0}~=~-\frac{1}{3}\big[ F_a(K^*K)+F_b(K^*K) \big] - \frac{1}{3}\big[ F_c(K^*K)+F_d(K^*K) \big], \label{K0K0VP} \\
&F_{\w\pi^0}~=~\big[F_a(\w\pi)+F_b(\w\pi)\big]+\frac{1}{18}\big[F_e(\w\pi)+F_f(\w\pi)\big], \\
&F_{\phi\pi^0}~=~\frac{\sqrt{2}}{18}\big[F_e(\phi\pi)+F_f(\phi\pi)\big].
\end{align}

The form factors for $e^+e^-\rightarrow V(T)\eta^{(}{'}^{)}$ are   mixtures  of the $\eta_q$ and $\eta_s$ components:
\begin{align}
&F_{V(T)\eta}=\cos\theta F_{V(T)\eta_q}-\sin\theta F_{V(T)\eta_s}, \\
&F_{V(T)\eta'}=\sin\theta F_{V(T)\eta_q}+\cos\theta F_{V(T)\eta_s},
\end{align}
where $V=\p^0,\w,\phi$ and
\begin{align}
&F_{\p^0\eta_q}~=~\big[ F_a(\p\eta_q)+F_b(\p\eta_q) \big]+\frac{5}{18}\big[ F_e(\p\eta_q)+F_f(\p\eta_q) \big], \\
&F_{\p^0\eta_s}~=~-\frac{\sqrt{2}}{6}\big[ F_e(\p\eta_s)+F_f(\p\eta_s) \big], \\
&F_{\w\eta_q}~=~\frac{1}{3}\big[ F_a(\w\eta_q)+F_b(\w\eta_q) \big]+\frac{5}{54}\big[ F_e(\w\eta_q)+F_f(\w\eta_q) \big], \\
&F_{\w\eta_s}~=~-\frac{\sqrt{2}}{18}\big[ F_e(\w\eta_s)+F_f(\w\eta_s) \big], \\
&F_{\phi\eta_q}~=~-\frac{5\sqrt{2}}{54}\big[ F_e(\phi\eta_q)+F_f(\phi\eta_q) \big], \\
&F_{\phi\eta_s}~=~-\frac{2}{3}\big[ F_a(\phi\eta_s)+F_b(\phi\eta_s) \big]-\frac{1}{27}\big[ F_e(\phi\eta_s)+F_f(\phi\eta_s) \big].
\end{align}

Similarly, based on Eq.(\ref{TPrelationsOfABCD}), form factors of the $e^+e^-\rightarrow TP$ channels can be written as:
\begin{align}
&F_{a_2^+\pi^-}~=~-F_{a_2^-\pi^+}~=~\big[ F_a(a_2\pi)+F_b(a_2\pi) \big], \\
&F_{K_2^{*+}K^-}~=~\frac{2}{3}\big[ F_a(K_2^*K)+F_b(K_2^*K) \big] - \frac{1}{3}\big[ F_c(K_2^*K)+F_d(K_2^*K) \big], \label{K+K-TP} \\
&F_{K_2^{*-}K^+}~=~-\frac{1}{3}\big[ F_a(K_2^*K)+F_b(K_2^*K) \big] + \frac{2}{3}\big[ F_c(K_2^*K)+F_d(K_2^*K) \big], \\
&F_{K_2^{*0}\bar{K}^0}~=~F_{\bar{K}_2^{*0}K^0}~=~-\frac{1}{3}\big[ F_a(K_2^*K)+F_b(K_2^*K) \big] - \frac{1}{3}\big[ F_c(K_2^*K)+F_d(K_2^*K) \big]. \label{K0K0TP}
\end{align}
The abbreviations  $a_2, K_2^*$ correspond to the tensor meson $a_2(1320)$ and $K_2^*(1430)$, respectively.

\section{Numerical Results and Discussions}

Using Eqs.(\ref{FaResult})-(\ref{FdResult}), and    other  input parameters, we can calculate cross sections for the processes $e^+e^-\rightarrow VP$ and $e^+e^-\rightarrow TP$. In Tab.~\ref{tab:CrossOfVP}, we have collected the results for cross sections at $\sqrt{s}=3.67$GeV, together with the experimental data from CLEO-c collaboration~\cite{Adam:2004pr,Adams:2005ks} (see Ref.~\cite{Ablikim:2004kv} for BES measurements), and the results at $\sqrt{s}=10.58$GeV, together with the data measured by Belle~\cite{Shen:2013okm} and Babar~\cite{Aubert:2006xw} collaborations. As we have discussed before,  three different types of transverse momentum distribution functions were used, denoted as $S1,S2$ and $S3$ respectively.  $S1$ denotes the calculation   without intrinsic transverse momentum distribution, $S2$ and $S3$ are obtained with the distributions in Eqs.(\ref{wf1}) and (\ref{wf2}), respectively.  Theoretical errors are obtained by varying   $\Lambda_{\rm QCD}=(0.25\pm0.05)$GeV, and the factorization scale $t$   from $0.75t$ to $1.25t$ (without changing $1/b_i$).


\begin{table}[htbp]
\caption{Cross sections of $e^+e^-\to VP, TP$ at $\sqrt s=3.67$ GeV and $\sqrt s=10.58$ GeV.   $S1$ denotes the calculation   without intrinsic transverse momentum distribution, $S2$ and $S3$ are obtained  with the distributions as Eqs.(\ref{wf1}) and (\ref{wf2}). The experimental measurements from Refs.~\cite{Adam:2004pr,Adams:2005ks,Shen:2013okm,Aubert:2006xw} are also shown. Theoretical errors are obtained by varying   $\Lambda_{\rm QCD}=(0.25\pm0.05)$GeV, and the factorization scale $t$   from $0.75t$ to $1.25t$ (without changing $1/b_i$).}
\label{tab:CrossOfVP}
\scalebox{0.8}{\begin{tabular}{c|cccc|cccc}
\hline\hline
  & \multicolumn{4}{c|}{ $\sqrt s=3.67$ GeV } & \multicolumn{4}{c}{ $\sqrt s=10.58$ GeV } \\
 Channel & $~~\sigma_{S1}$(pb)~~ & $\sigma_{S2}$(pb)~~& $\sigma_{S3}$(pb)~~ & $\sigma_{\mathrm{exp}}$(pb)~~ &~~ $\sigma_{S1}$(fb)~~ & $\sigma_{S2}$(fb)~~ & $\sigma_{S3}$(fb)~~ & $\sigma_{\mathrm{exp}}$(fb)\\
\hline
$\p^\pm\pi^\mp$           & $6.80\pm1.18$   & $3.38\pm0.53$  & $3.95\pm0.63$    & $4.8_{-1.2-0.5}^{+1.5+0.5}$  & $0.66\pm0.10$  & $0.53\pm0.08$ & $0.60\pm0.09$ & $$   \\
$\p^0\pi^0$           & $3.38\pm0.60$   & $1.69\pm0.27$  & $1.99\pm0.32$    & $3.1_{-1.2-0.4}^{+1.0+0.4}$  & $0.25\pm0.05$  & $0.20\pm0.04$ & $0.23\pm0.04$ & $$   \\
$K^{*\pm} K^\mp$       & $10.13\pm0.91$  & $5.27\pm0.50$  & $5.39\pm0.35$    & $1.0_{-0.7-0.5}^{+1.1+0.5}$  & $1.15\pm0.10$  & $0.94\pm0.08$ & $1.02\pm0.08$ & $0.18_{-0.12}^{+0.14}\pm0.02$   \\
$K^{*0}\bar{K}^0+ \overline K^{*0} {K}^0$ & $61.94\pm13.76$  & $31.34\pm6.15$ & $31.85\pm6.25$   & $23.5_{-3.9-3.1}^{+4.6+3.1}$ & $6.65\pm1.20$  & $5.39\pm0.93$ & $5.88\pm1.02$ & $7.48\pm0.67\pm0.51$   \\
$\w\pi^0$             & $24.94\pm4.59$  & $12.41\pm2.08$ & $15.18\pm2.59$   & $15.2_{-2.4-1.5}^{+2.8+1.5}$ & $2.38\pm0.40$  & $1.90\pm0.31$ & $2.16\pm0.35$ & $$  \\
$\phi\pi^0$           & $1.2\times10^{-4}$ & $1.2\times10^{-4}$ & $1.2\times10^{-4}$ & $<2.2$     & $2.2\times10^{-3}$ & $2.2\times10^{-3}$ & $2.2\times10^{-3}$    & $$  \\
\hline
$\p^0\eta$            & $14.37\pm2.10$  & $7.21\pm0.96$  & $8.10\pm1.06$   &  $10.0_{-1.9-1.0}^{+2.2+1.0}$ & $1.10\pm0.13$  & $0.89\pm0.11$ & $1.03\pm0.12$ & \\
$\p^0\eta'$           & $8.22\pm1.19$   & $4.10\pm0.54$  & $4.57\pm0.59$   &  $2.1_{-1.6-0.2}^{+4.7+0.2}$  & $1.03\pm0.11$  & $0.83\pm0.09$ & $0.93\pm0.10$ & \\
$\w\eta$              & $1.31\pm0.20$   & $0.65\pm0.09$  & $0.77\pm0.11$   &  $2.3_{-1.0-0.5}^{+1.8+0.5}$  & $0.10\pm0.01$  & $0.081\pm0.011$ & $0.094\pm0.012$ & \\
$\w\eta'$             & $0.75\pm0.11$   & $0.37\pm0.05$  & $0.43\pm0.06$   &  $<17.1$                      & $0.094\pm0.011$ & $0.076\pm0.009$ & $0.086\pm0.010$ & \\
$\phi\eta$            & $17.82\pm3.34$  & $9.21\pm1.51$  & $8.23\pm1.32$   &  $2.1_{-1.2-0.2}^{+1.9+0.2}$  & $2.11\pm0.30$  & $1.75\pm0.23$ & $1.84\pm0.25$ & $2.9\pm0.5\pm0.1$\\
$\phi\eta'$           & $21.97\pm4.13$  & $11.36\pm1.87$ & $10.20\pm1.65$  &  $<12.6$                      & $2.81\pm0.42$  & $2.31\pm0.33$ & $2.47\pm0.35$ & \\
\hline\hline
\hline
$a_2^\pm\pi^\mp$             & $43.88\pm13.98$       & $20.34\pm6.59$       & $28.96\pm8.62$     & $$ & $6.66\pm1.73$       & $4.96\pm1.30$        & $6.06\pm1.58$        & $$   \\
$K_2^{*\pm}K^\mp$       & $60.57\pm15.89$       & $27.81\pm7.45$       & $33.81\pm8.98$     & $$ & $11.48\pm2.45$      & $8.48\pm1.79$        & $9.98\pm2.15$       & $8.36\pm0.95\pm0.62$   \\
$K_2^{*0}\bar{K}^0+ \overline K_2^{*0} {K}^0$ & $3.2\times10^{-2}$    & $1.1\times10^{-2}$   & $1.3\times10^{-2}$ & $$ & $8.8\times10^{-3}$  & $6.0\times10^{-3}$   & $7.3\times10^{-3}$ & $1.65^{+0.86}_{-0.78}\pm{0.27}$   \\
\hline
\end{tabular}}
\end{table}

A few remarks are in order.
\begin{itemize}
\item
 Results at different center of mass energy $\sqrt{s}$ can be used to study  the $1/s^n$ dependence of cross sections.  From our results at $\sqrt{s}=3.67$GeV and $10.58$GeV,  the averaged value  is about  $n=4.1$ for $e^+e^-\rightarrow VP$ and $n=3.9$ for $e^+e^-\rightarrow TP$~\footnote{We correct here the improper statement in Ref.~\cite{Lu:2007hr}.}. This favors the $1/s^4$ scaling, which is consistent with the constituent scaling rule~\cite{Lepage:1980fj,Brodsky:1981kj}.  The fitted result from experimental data  is  $n=3.83\pm0.07$ and $3.75\pm0.12$ for $e^+e^-\rightarrow K^*(892)^0\bar{K}^0$ and $\w\pi^0$, respectively~\cite{Shen:2013okm}.

\item
From Table \ref{tab:CrossOfVP}, we can see that,   cross sections for many processes are large enough to be measured,  such as the   $e^+e^-\rightarrow \p\pi,\p\eta,\w\pi$ and $a_2^\pm\pi^\mp$ at $\sqrt{s}=10.58$GeV, and $e^+e^-\rightarrow a_2^\pm\pi^\mp,K_2^{*\pm}K^{\mp}$ at $\sqrt{s}=3.67$GeV. We suggest the experimentalists to measure these channels especially at BESIII~\cite{Asner:2008nq} and Belle-II in future.

\item For the channels $e^+e^-\rightarrow K^{*\pm}K^{\mp}$, there are very poor measurements from CLEO collaboration \cite{Adams:2005ks}, since the charged $K^*$ meson is reconstructed by    three--body decays:   $K^{*\pm}\rightarrow K^0\pi^{\pm}\rightarrow3\pi$, with large   systematic uncertainties.   Our results are larger than the central of experimental data. We hope the future experimental measurements can clarify this difference more clearly.

\item If we neglect the  photon-enhanced amplitudes $F_{e,f}$, and assume the flavor SU(3) symmetry, one has the relations for cross sections: $\sigma(\omega\pi^0): \sigma(\rho^\pm\pi^\mp):\sigma(\rho^0\pi^0): \sigma(K^{*\pm}K^\mp): \sigma(K^{*0}\bar K^0 + \overline K^{*0}K^0)$ = $1:2/9:1/9: 2/9: 8/9$.

\item At $\sqrt{s}=3.67$ GeV, we have  $\sigma(e^+e^-\to \rho^\pm\pi^\mp)= 2 \sigma(e^+e^-\to \rho^0\pi^0)$, while the photon-enhanced contribution becomes more important at $\sqrt{s}=10.58 $GeV, and the ratio $\sigma(e^+e^-\to \rho^\pm\pi^\mp)/\sigma(e^+e^-\to \rho^0\pi^0)$ is approximately 2.5.

\item In the SU(3) limit, we expect $\sigma(\omega\pi^0)/\sigma(K^{*0}\bar K^0 + \overline K^{*0}K^0)=9/8>1$, however our calculation has indicated that the cross section $\sigma(\omega\pi^0)$ is smaller than that for $e^+e^-\to K^{*0}\bar K^0 + \overline K^{*0}K^0$ by a factor of 2 to 3. One reason arises from the fact that the  decay constants $f_\pi f_{\omega}$ is about $30\%$ smaller than  $f_K f_{K^*}$.   The chiral scale parameter $m_{0}^K$ will further enhance the cross sections.

\item
On the experiment side, the ratios $R_{VP}$ and $R_{TP}$ are introduced to explore the SU(3) symmetry breaking effect in the $e^+e^-\rightarrow K^*K$ and $e^+e^-\rightarrow K_2^*K$ processes, with the definition
\begin{align}
R_{VP}=\frac{\o(e^+e^-\rightarrow K^*(892)^0\bar{K}^0)}{\o(e^+e^-\rightarrow K^*(892)^-K^+)},~~~R_{TP}=\frac{\o(e^+e^-\rightarrow K_2^*(1430)^0\bar{K}^0)}{\o(e^+e^-\rightarrow K_2^*(1430)^-K^+)}.
\end{align}
In  the PQCD framework,   this ratio can be written as
\begin{align}
R=\bigg|\frac{(F_a+F_b)+(F_c+F_d)}{2(F_a+F_b)-(F_c+F_d)}\bigg|^2=\bigg|\frac{1+\frac{F_c+F_d}{F_a+F_b}}{2-\frac{F_c+F_d}{F_a+F_b}}\bigg|^2.
\end{align}
In SU(3) symmetry limit, the wave functions of $K,K^*$ and $K_2^*$ is   symmetric or antisymmetric under the exchange of the momentum fractions of quark and antiquark, and thus  the relations in Eq.(\ref{VPrelationsOfABCD})  are obtained. Then one can  drive   $R_{VP}=4$. One source  of the SU(3) symmetry breaking   is that the $s$ quark is heavier than $q(=u,d)$ quark and carries more momentum in the final state meson, therefore the gluon which generates $\bar{s}s$ is harder than the $\bar{q}q$ one. In this case, the coupling constant in the $\bar{s}s$ process is smaller. Consequently, the amplitude  $|F_a+F_b|$ will be  smaller than $|F_c+F_d|$, and thus $R_{VP}$ is expected larger than 4.

From Table \ref{tab:CrossOfVP}, one can obtain theoretical  results for  $R_{VP}$:
\begin{gather}
R_{VP}(\sqrt{s}=3.67{\rm GeV})\simeq5.99, ~~~
R_{VP}(\sqrt{s}=10.58 {\rm GeV})\simeq5.76.
\end{gather}

\item
At  $\sqrt{s}=3.67$ GeV, the CLEO-c collaboration  \cite{Adams:2005ks} has measured the ratio:
\begin{align}
R_{VP}^{Exp}(\sqrt{s}=3.67{\rm GeV})=23.5_{-26.1}^{+17.1}\pm12.2,
\end{align}
with very large error-bar. Its central value is significantly larger, but within the errors  it is    consistent with our theoretical results.
Belle collaboration gives the results at $\sqrt{s}=10.52$GeV, $10.58$GeV and $10.876$GeV, respectively~\cite{Shen:2013okm}
\begin{align}
&R_{VP}^{Exp}~>~4.3,~~~20.0,~~~5.4.
\end{align}
Note that  in the region near $10.58$GeV,   Belle result is significantly larger than our expectation, which might come from the  $\Upsilon(4S)$ resonance contribution. Off the $\Upsilon(4S)$ resonance, the experimental results are consistent with our theoretical calculations.

\item Due to the charge conjugation invariance, we have the relations for the $e^+e^-\to TP$ transition amplitude given in Eq.~\eqref{TPrelationsOfABCD}. Thus only three channels are allowed: $e^+e^-\to a_2^\pm\pi^\mp$, $e^+e^-\to K_2^{*\pm}K^{\mp}$ and $e^+e^-\to K_2^{*0}\bar K^0+ \overline K_2^{*0}  K^0$.

If one further assume V-spin symmetry, the process $e^+e^-\rightarrow K_2^{*0}\bar{K}^0+ \overline K_2^{*0} {K}^0$ is highly suppressed since $F_a+F_b\sim -(F_c+F_d)$.  From Table \ref{tab:CrossOfVP}, one can obtain theoretical  results for    $R_{TP}$:
\begin{gather}
R_{TP}\lesssim10^{-4}.
\end{gather}
This is consistent with the Belle data~~\cite{Shen:2013okm}:
\begin{eqnarray}
R_{TP}^{Exp}~<~1.1,~~~0.4,~~~0.6.
\end{eqnarray}

\item

The theoretical uncertainties in our calculation are mainly from the uncertainties of the meson wave functions. The   longitudinal distribution amplitudes in  exclusive B decays  will give about $10\%-20\%$ uncertainties~\cite{Kurimoto:2006iv}. When  the transverse momentum distribution functions are  introduced in Eqs.(\ref{wf1}) and (\ref{wf2}), the contribution from the large-$b$ region will be suppressed. This suppression makes the PQCD approach more self-consistent. Comparing the different results in Table \ref{tab:CrossOfVP}, one  can observe   severe suppressions  especially at  $\sqrt{s}=3.67$GeV: the suppression is about $50\%$ for $S2$ and about $40\%$ for $S3$. Since the results depend on the explicit form of transverse momentum distribution, more accurate transverse momentum dependent wave functions and more experimental results would be valuable.

\item
In this calculation, we have  limited ourselves  to the  leading-order accuracy. The next-to-leading order (NLO) calculation  is complicated~\cite{Li:2012nk,Cheng:2014gba,Cheng:2014fwa} that will be presented in a future publication. As an  estimation of the size of the NLO contribution, we vary  $\Lambda_{\rm QCD}$ and the factorization scale $t$ in Eq.(\ref{hst1}): $\Lambda_{\rm QCD}=(0.25\pm0.05)$GeV, and changing the hard scale $t$ from $0.75t$ to $1.25t$ (without changing $1/b_i$). We find that our results are not sensitive to these variations. It implies that the NLO contributions   are presumably  not very large.

\end{itemize}

\section{Conclusion}

Hard exclusive processes $e^+e^-\rightarrow VP$ and $e^+e^-\rightarrow TP$ at center of mass energy $\sqrt{s}=3.67$GeV and $10.58$GeV are investigated in the perturbative QCD framework  in this work. For the wave functions of the light mesons involved in the factorization amplitudes, we have employed various models of transverse momentum dependence of wave functions. At the center  of mass energy $\sqrt{s}=3.67$GeV,   two different transverse momentum distribution functions can give about $50\%$ and $40\%$ suppressions, respectively. The value $R_{VP}$ and $R_{TP}$ obtained from our results are consistent with the  experimental data. We found  that our theoretical results favor the $1/s^4$ scaling law for the cross sections. Most of our results are consistent with the experimental data and the others can be tested at the ongoing  BESIII  and forthcoming Belle-ILL experiments.

\section*{Acknowledgements}

The authors are grateful to Jian-Ping Dai, Hsiang-nan Li, Cheng-Ping Shen and Yu-Ming Wang for valuable  discussions. This work
is supported in part by National Natural Science Foundation of China
under Grant No.11575110, 11521505, 11655002, 11621131001, 11735010, Natural Science Foundation of Shanghai
under Grant No.~15DZ2272100,  by Shanghai Key Laboratory for Particle Physics and Cosmology,  and by Key Laboratory
for Particle Physics, Astrophysics and Cosmology, Ministry of Education.

\end{document}